# Surface-Confined Two-Dimensional Crystal Growth on a Monolayer

Yanyu Jia[1*,#], Fang Yuan[2,#], Guangming Cheng[3], Yue Tang[1], Guo Yu[1,4], Tiancheng Song[1], Pengjie Wang[1], Ratnadwip Singha[2], Ayelet July Uzan[1], Michael Onyszczak[1], Kenji Watanabe[5], Takashi Taniguchi[6], Nan Yao[3], Leslie M Schoop[2*], Sanfeng Wu[1*]

[1] Department of Physics, Princeton University, Princeton, New Jersey 08544, USA
[2] Department of Chemistry, Princeton University, Princeton, New Jersey 08544, USA
[3] Princeton Materials Institute, Princeton University, Princeton, New Jersey 08544, USA
[4] Department of Electrical and Computer Engineering, Princeton University, Princeton, New Jersey 08544, USA
[5] Research Center for Electronic and Optical Materials, National Institute for Materials Science, 1-1 Namiki, Tsukuba 305-0044, Japan
[6] Research Center for Materials Nanoarchitectonics, National Institute for Materials Science, 1-1 Namiki, Tsukuba 305-0044, Japan[#] These authors contributed equally to this work

[*] Email: sanfengw@princeton.edu, lschoop@princeton.edu, yanyuj@princeton.edu

## Abstract

**Conventional vapor deposition or epitaxial growth of two-dimensional (2D) materials and heterostructures is conducted in a large chamber in which masses transport from the source to the substrate. Here we report a chamber-free, on-chip approach for growing a 2D crystalline structures directly in a nanoscale surface-confined 2D space. The method is based on a surprising discovery of a rapid, long-distance, non-Fickian transport of a uniform layer of atomically thin palladium (Pd) on a monolayer crystal of tungsten ditelluride ($WTe_2$), at temperatures well below the known melting points of all materials involved. The resulting nanoconfined growth realizes a controlled formation of a stable new 2D crystalline material, $Pd_7WTe_2$ , when the monolayer seed is either free-standing or fully encapsulated in a van der Waals stack. The approach is generalizable and highly compatible with nanodevice fabrication, promising to expand the library of 2D materials and their functionalities.**

## Main

The development of 2D quantum materials and structures[1–8] is providing strong new impetus to a variety of fields in physics, engineering, and chemistry. Based on known layered bulk compounds, a computational survey[9] has indicated that there are more than 1, 000 layered van der Waals (vdW) crystals that can potentially be created down to the monolayer limit. Indeed, extensive efforts have been devoted to creating and optimizing 2D crystals in this library, especially e.g., graphene, boron nitride, transition metal dichalcogenides and many other monolayer crystals. Widely applied methods[1–8] include mechanical exfoliation, liquid exfoliation and direct crystal growth approaches such as chemical or physical vapor deposition and molecular beam epitaxy. 2D materials beyond this library, especially those without known 3D counterparts, remain largely unexplored. New approaches for 2D synthesis and characterizations are needed.

In general, the growth of 2D crystals involves transport of atoms or molecules from the source to the growth area on the substrate. This mass transport process typically occurs in an open space, e.g., in a furnace with noble gas carriers for vapor deposition or in a vacuum chamber for molecular beam epitaxy**.** Alternatively, a liquid metal environment has been employed for synthesizing 2D metal oxides[10]. Recently, a flux-assisted method has also been developed to grow many 2D crystals[11]. Inside a solid state environment, mass transport is expected to occur slowly via substitutional or interstitial atomic migration, following Fick's diffusion laws[12], which therefore does not provide an ideal transport mechanism for growing large area new materials. For instance, recent studies of reactions between metal and transition metal dichalcogenides have observed the formation of new materials only at interfacial area or near its very vicinity within ~ 100 nm extended region[13–15].

Nevertheless, mass transport phenomena and mechanisms through nanoconfined solid structures or surfaces remain a largely unexplored frontier, despite its critical importance to a wide range of fields including molecular biology, nanotechnology, and quantum material science. For instance, studies of ion and molecular transport through a carbon nanotube[16,17] or an atomically thin nano-slit formed inside a vdW structure[18,19] has raised many intriguing questions regarding nanofluidics[20,21]. Interesting possibilities, such as nearly frictionless flow[18,22], critical wetting[23], nanoconfined chemistry[24] on a 2D crystal, and collective phases of atoms adsorbed on a monolayer[25], to name a few, await experimental explorations. In this work, we report a surprising observation of rapid, non- Fickian mass transport of atomically thin metals on monolayer crystals over a very long distance at temperatures well below the melting points of all materials. As an example, we demonstrate in detail the propagation process and results of 2D Pd film transport on monolayer $WTe_2$. The experiments establish a novel and generalizable approach, strictly confined in a 2D nano-space near the surface of a monolayer crystal, for the growth of new 2D materials and crystalline structures inaccessible with preexisting methods.

## Results

### *Long-distance mass transport on a monolayer*

Our experiments start with fabricating a vdW stack (Fig. 1a), in which monolayer $WTe_2$ and pre-deposited Pd islands (~ 20 nm thick) are fully encapsulated by the chemically inert hexagonal boron nitride (hBN). The monolayer $WTe_2$ is mechanically exfoliated from its bulk[26–28], and the Pd islands are created on the bottom hBN substrate using nanolithography and metal deposition techniques. The fully encapsulated vdW stacks are created and placed on top of a 285 nm $SiO_2$/Si substrate using the standard 2D dry transfer techniques in an argon-filled glovebox (see Methods, Fig. 1 and Extended Data Fig. 1). During the transfer process, the entire stack is heated up to ~ 170 °C, at which temperature ($T$) we observe no visible mass transport phenomena under an optical microscope (Fig. 1b). Upon heating the vdW stack to above 190 °C, we observe a clear expansion of Pd atoms seeded from the Pd islands underneath $WTe_2$ (see Figs. 1c & d for cartoon illustrations). This is indicated by the optically darker region seen in Fig. 1e, in which we show a series of optical microscope images that reveal a continuous, rapid expansion of the dark region. Holding $T$ at ~ 200°C for ~ 70 minutes, we find that the expansion nearly covers the entire monolayer $WTe_2$ (Figs. 1f & g). In Supplementary Video 1, we record the continuous mass-spreading process of device D1. Extended Data Fig. 2 shows images of the same device taken under an atomic force microscope (AFM). It clearly reveals that, after expansion, the seed Pd islands become thinner and narrower. Near the separated Pd islands (i.e., not in contact with monolayer $WTe_2$), no visible change of any material is seen. The expansion stops at the monolayer edge, i.e., no spreading of any material is seen on hBN in the absence of $WTe_2$.

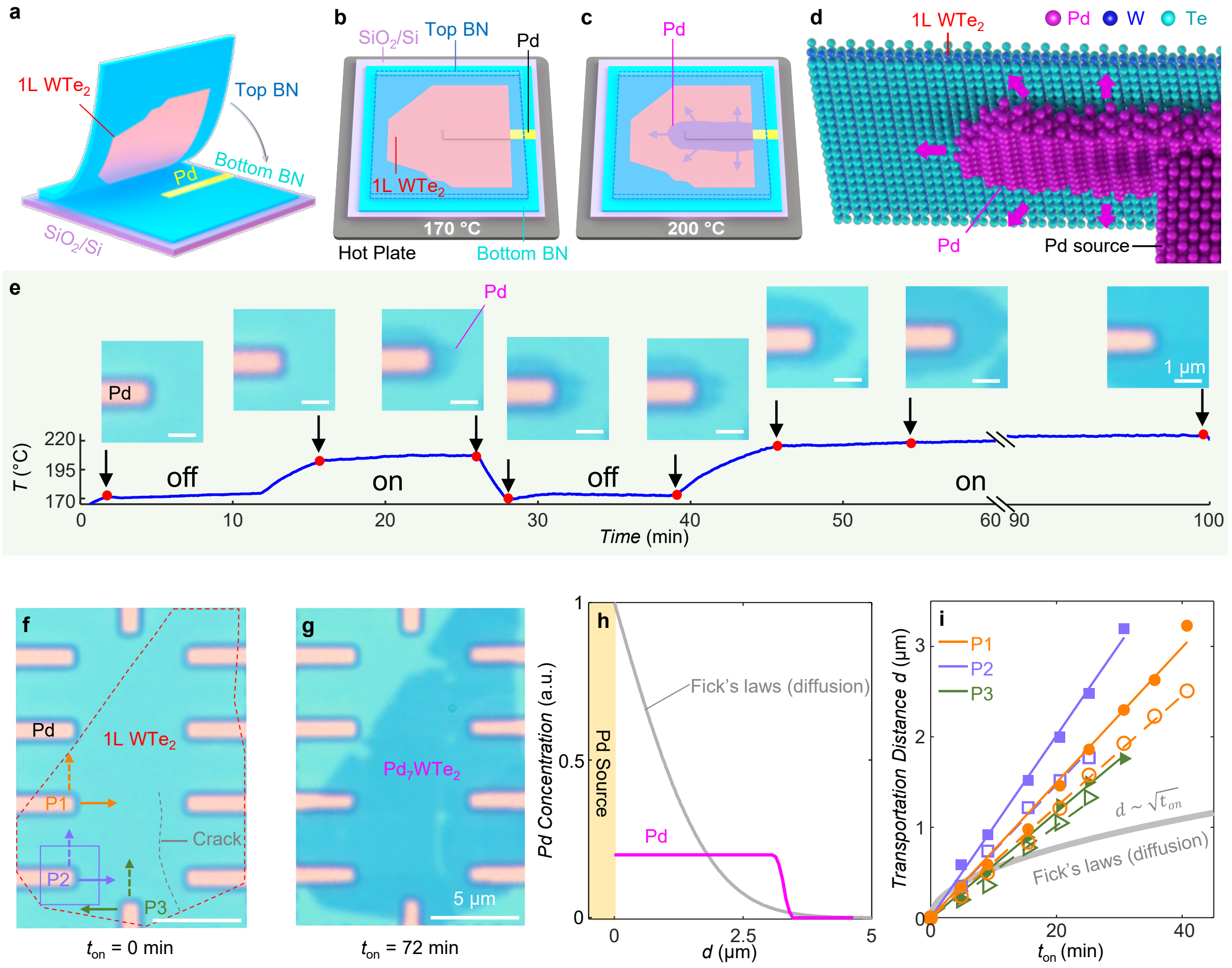

**Figure 1 | Long-distance anomalous mass transport of Pd atoms on monolayer $WTe_2$. a,** Cartoon illustration of the vdW stacked structure prepared for the experiment. **b & c,** Cartoon illustrations of the observed mass transport phenomena on monolayer (1L) $WTe_2$**.** At 170 °C, no visible change of any materials is observed (**b**), while at 200 °C, clear mass spreading is seen (**c**). **d,** Cartoon illustration of the Pd transport process on monolayer $WTe_2$ crystal. **e,** Controlled mass transport, by modifying the temperature. The optical images are taken throughout the process near one Pd island in device D1 (indicated by the purple square in **f**). The transport can be completely turned on/off by changing the temperature. **f,** Optical image of device D1 prior to Pd transport. Monolayer $WTe_2$, invisible in this image, is outlined. $t_{on}$ is the time for the device being held at a temperature above 190°C during the process in **e**. **g,** Optical image of the same device after holding the temperature at ~ 200 °C for ~ 70 minutes. The product 2D crystal $Pd_7WTe_2$ is now visible optically. **h,** Illustration of the contrast between the observed concentration profile of D1 (purple curve, based on optical contrast) and the expectation from the Fick's laws (grey curve). **i**, Time dependent transport distance $d(t_{on})$, extracted from the images taken from device D1. The data points, with different shapes and colors, correspond to the transport traces indicated by the colored arrows in **f**, respectively.

We highlight the key features of this utterly unexpected observation. In the standard diffusion process dictated by the Fick's laws, the diffusion flux is determined by the concentration gradient, which leads to[12] (*i*) an exponential decay in the spatial profile of concentration, $c(x)$, of the diffusive atoms, i.e., $c(x) \sim \exp(-x^2/4Dt)$, and (*ii*) a time-dependent diffusion distance $d(t) \sim \sqrt{t}$. Here $D$ is the diffusion coefficient and ($x$, $t$) are the space time coordinates. Both descriptions are invalid in our observations. The spatial distribution of the propagating Pd, as already indicated by the uniform optical contrast in Figs. 1e & g, appears to be of a constant concentration away from the seed Pd and develops a sharp drop to zero at the front. This is in stark conflict with the Fickian diffusion, as illustrated in Fig. 1h. The thickness of the uniform Pd ultrathin film spread on monolayer $WTe_2$ is ~ 0.8 nm, measured from AFM (Extended Data Fig. 3). In Fig. 1i, we further extract the time-dependent transport distance, measured between the moving front and the Pd source along two directions based on 3 sources, P1-P3, as indicated in Fig. 1f. We find a linear time dependence, i.e., $d(t) \sim t$. Namely, the front propagates at a constant speed, roughly on the order of ~ 0.1 μm/min. These features uncover that the unusual mass transport phenomenon observed here is non-Fickian, i.e., non-diffusive. We note that while a similar device geometry has been previously used to study the diffusive and interfacial reactions between Pd and multilayer $WTe_2$[13,14], our finding of the non-Fickian rapid mass transport is distinct qualitatively; here the Pd transport distance is very long and only limited by the size of the seed monolayer, critical for enabling large area 2D growth.

Indeed, the entire process occurs as if the Pd atoms are flowing on monolayer $WTe_2$, like a liquid! The challenge in this picture is that the melting point of Pd metal is 1555 °C. We don't observe any obvious melting of the Pd bulk islands at the temperatures applied here. Nevertheless, the fact that the Pd transport can be completely turned on and off (Fig. 1e) by slightly modifying $T$ suggests the presence of a critical temperature (~ 190 °C). This may be regarded as a clue that certain structural phase transition still occurs. Interestingly, it is known that the melting point of many metals with a nanoscale size can be significantly depressed compared to their bulk value[29]. We thus suspect that the very surface layer of Pd, which directly contacts $WTe_2$, might melt or undergo a structural transition. Another important aspect regards the thickness of the spreading Pd film, which appears to be highly uniform (~ 0.8 nm) and is the same in all samples (Extended Data Fig. 3). At this stage, we don't know what determines this exact thickness but would like to mention that the interesting theoretical idea of "critical wetting" on a monolayer crystal[23] may be considered. It is possible that, above 190°C, we have realized an atomically thin 2D liquid metal (perhaps the partial Pd layer), a highly unusual form of matter not realized previously. Future efforts that combine experiments and theories are necessary to fully understand the exact mechanism of such anomalous non-Fickian mass transport. Importantly, the experimental fact observed here immediately implies new possibilities for growing 2D materials.

## *Surface-nanoconfined Two-dimensional growth on a monolayer*

The direct chemical consequence of the Pd transport is the growth of a new 2D material converted from the precursor monolayer (i.e., $WTe_2$). We perform high-resolution scanning transmission electron microscope (STEM) studies on the vdW stack to uncover its atomic structure (see Methods). All STEM data is taken at room temperature, under which condition the material structure is in a solid phase and stable (i.e., no active Pd transport). Fig. 2a presents a bright-field STEM image of the stack (cross-section view) at a selected location near the Pd front, where we see both the Pd-covered regime and the bare monolayer $WTe_2$ (Fig. 2a inset). The new material is clearly thicker, consistent with the darker optical appearance in Fig. 1 and the AFM measurements (Extended Data Fig. 3). The image immediately confirms the extremely sharp Pd front, with only ~ 5 nm lateral interface at the transition (Fig. 2a). Energy-dispersive x-ray spectroscopy (EDX) reveals the dominant presence of Pd atoms of the new structure on the left side, while no Pd is found on the right (Figs. 2b & c). W and Te atoms are present on both sides, with a composition ratio of 1:2, as expected. In the new material region shown in Fig. 2b, the atomic ratio of Pd:W:Te is found to be 68.2%:10.7%:21.1%, very close to 7:1:2. In Extended Data Figs. 4 & 5, we present a systematic EDX analysis on different locations. All yield a similar atomic ratio close to 7:1:2. We hence attribute this new 2D material structure to a nominal chemical formula $Pd_7WTe_2$.

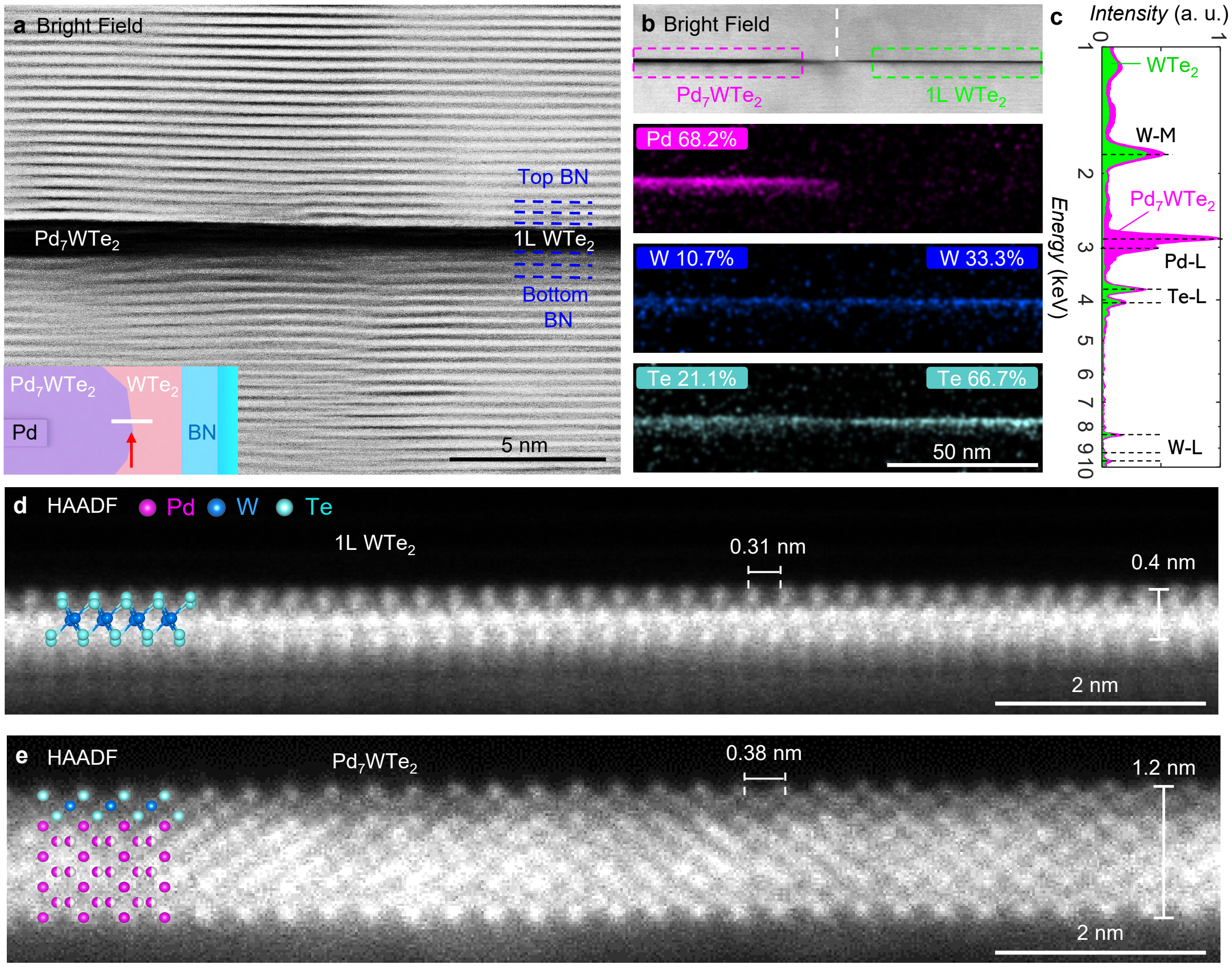

**Figure 2 | Nanoconfined crystal growth on monolayer $WTe_2$. a,** A bright-field STEM image (cross-section) obtained at the $Pd_7WTe_2$ and $WTe_2$ interface in a vdW stacked device (D2). The position is indicated by the cartoon inset. **b,** EDX analysis on the atomic ratio of both $Pd_7WTe_2$ (left side) and monolayer $WTe_2$ (right side) in D3. **c.** EDX spectra of both sides, on areas indicated by the dashed box in **b**. W-M, W-L and Te-L energy peaks are well resolved in both areas, while significant Pd-L peaks only emerge in $Pd_7WTe_2$ region. **d.** A HAADF STEM image of monolayer $WTe_2$ in the device D3. **e.** A HAADF STEM image of the new 2D crystalline structure $Pd_7WTe_2$ synthesized on the same device. A proposed crystal structure is superimposed (see Extended Data Fig. 6 for details).

Remarkably, this new material exhibits a crystalline structure, as revealed in the high-angle annular dark field (HAADF) STEM measurements. We first show the HAADF image taken on a pristine $WTe_2$ monolayer region (Fig. 2d) in the same sample (cross-section), confirming the expected lattice structure. Fig. 2e presents the same HAADF STEM measurements on the $Pd_7WTe_2$ region, which interestingly uncovers a crystalline structure, a qualitatively different one from the neighboring $WTe_2$ monolayer. The new structure, dominated by Pd atoms, arranges compactly in 2D form. We suspect that the W and Te atoms remain bonded and are located on the top of the stack, as evidenced by the slightly darker top three layers in Fig. 2e. Our experiments, unfortunately, are not able to distinguish the species of atoms site by site, and hence its confirmation requires future studies. A lateral interatomic spacing of ~ 0.38 nm is observed from this side view, noticeably larger than the 0.31 nm observed in the pristine $WTe_2$, which indicates a structural change of $WTe_2$ inside $Pd_7WTe_2$. In Extended Data Fig. 6, we compare the measured distances with those observed in elemental bulk Pd; clear differences are identified. We note that the process here is distinguished from vdW epitaxy[30] as we do not observe a van der Waals gap between Pd layer and the W-Te layer.

To visualize the in-plane atomic structure of this new 2D crystal, we prepared another sample on a standard TEM grid (see Methods), where the monolayer is fully suspended on holes. The Pd is pre-deposited on the TEM grid frame (Fig. 3a) and no hBN is used to ensure the exposure of the target crystal. By the same heating procedure, Pd transport is introduced on the suspended $WTe_2$ and forms the new crystal. The experiments confirm that the mass transport happens between Pd and $WTe_2$ and that the hBN is not essential (although the use of hBN stacks may improve the quality of the new crystal). In Fig. 3b, we present a high-resolution HAADF STEM image, in a plan-view, at the boundary between monolayer $WTe_2$ (left) and $Pd_7WTe_2$ (right). Both materials can be resolved with atomic resolution. The direction of the zig-zag W chain of monolayer $WTe_2$ is labeled. Fast Fourier transformation (FFT) of the monolayer $WTe_2$ area yields its characteristic diffraction pattern (Fig. 3c), consistent with its rectangular unit cell and strong lattice anisotropy. The high-resolution STEM image of the $Pd_7WTe_2$ regime uncovers a qualitatively different crystal structure, whose FFT pattern, shown in Fig. 3d, displays a six-fold symmetric pattern that is absent in $WTe_2$. Indeed, a zoom-in view of the $Pd_7WTe_2$ lattice shows an apparent hexagonal lattice (Fig. 3e). The characteristic interatomic length of 0.38 nm observed in the cross-section image (Fig. 2e) can also be identified here (Fig. 3e). We have further performed EDX analysis on such suspended TEM devices, which again yields an atomic composition of Pd:W:Te = 7:1:2, the same as the hBN-encapsulated samples (Fig. 3f).

These observations also confirm that the same crystalline structure of $Pd_7WTe_2$ is obtained for the two growths (hBN-encapsulated growth *v.s.* suspended growth). Specifically, both samples were fabricated on monolayer $WTe_2$, i.e., the W and Te atoms are of the same density; hence the EDX result of the same atomic composition confirms that the 2D Pd density is the same for both growths. Pd on both films shows a crystalline structure, in which we find an identical lattice constant (i.e., the 0.38 nm inter-atomic distance). Based on these observations, we propose a unique atomic structural model for the new material $Pd_7WTe_2$ (see Methods and Extended Data Fig. 6). As far as we know, neither the unusual hexagonal form of the 2D Pd layer nor a crystalline $Pd_7WTe_2$ structure has been reported prior to this work.

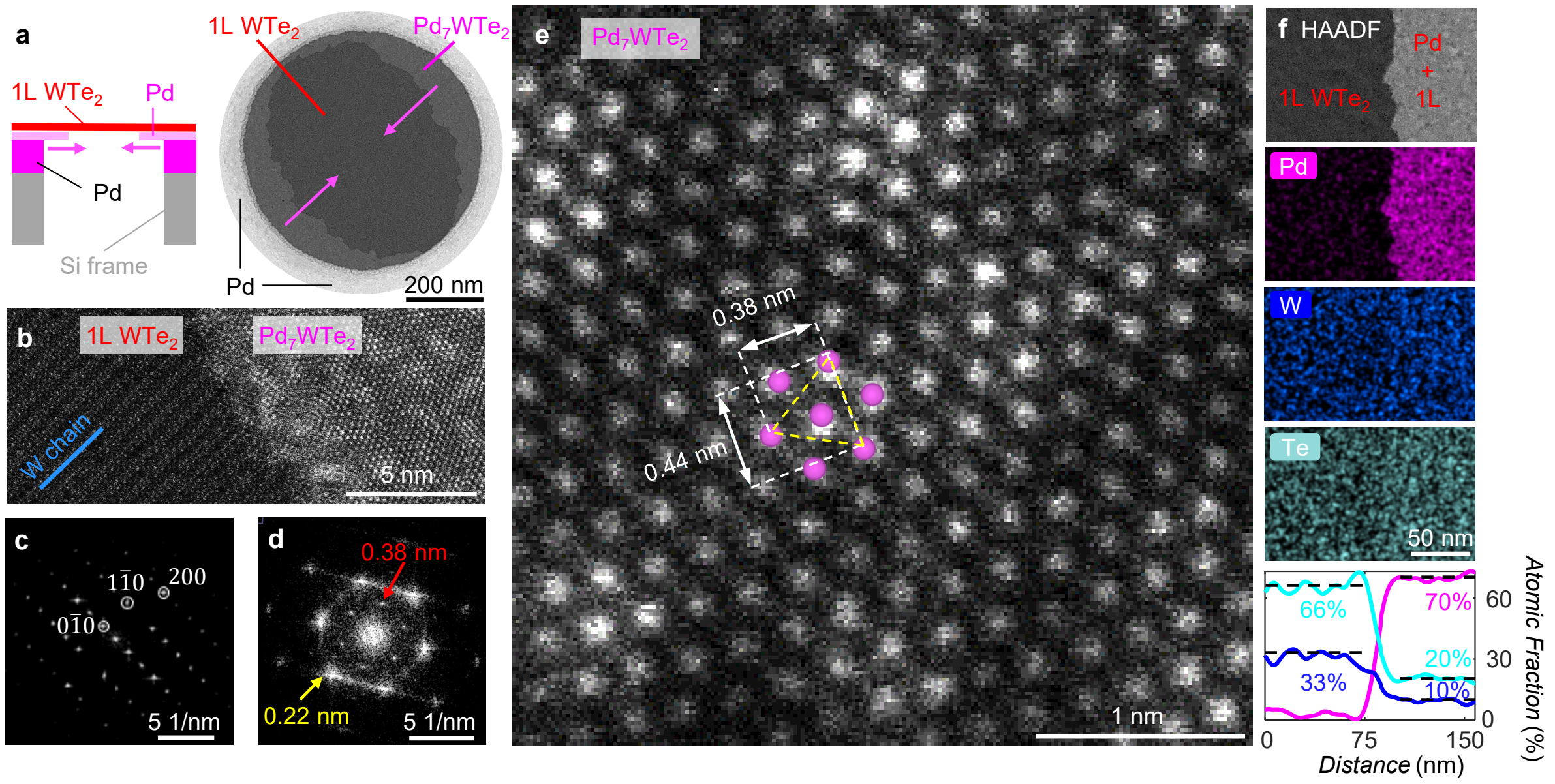

**Figure 3 | In-plane lattice structure of $Pd_7WTe_2$. a,** A STEM image of a suspended TEM device D4 (right), with a cartoon illustration of its structure (left). Pd is coated on the grid, after which monolayer $WTe_2$ is placed on top. The Pd transport is introduced upon heating the sample above 190 °C and the new 2D material $Pd_7WTe_2$ is formed on the suspended $WTe_2$. Regions of different materials are indicated. **b,** An atomic-resolution STEM image of a region interfacing $Pd_7WTe_2$ and monolayer $WTe_2$. The W chain in $WTe_2$ is denoted. **c.** An FFT pattern of the monolayer $WTe_2$ crystal with its Miller indices indicated. **d.** An FFT pattern of the $Pd_7WTe_2$ crystal with six-fold symmetry. The characteristic periodicity 0.38 nm and 0.22 nm are indicated. **e**. A magnified STEM image from $Pd_7WTe_2$ crystal, revealing its apparent hexagonal lattice**. f.** Elemental mappings and line-profile composition analysis on the suspended TEM sample near the interface between monolayer $WTe_2$ and $Pd_7WTe_2$.

We further note that the as-grown $Pd_7WTe_2$ is highly uniform over the entire area defined by the seed monolayer $WTe_2$. In Extended Data Fig. 7, we confirm the thickness uniformity over a large area (~ microns) directly based on atomic scale STEM measurements. In Extended Data Fig. 8, we show that the lattice orientation of $Pd_7WTe_2$ is determined by the seed monolayer $WTe_2$, as revealed by the STEM analysis on various front locations on a single monolayer, separated by large distances. Namely, the crystal orientation of as-grown $Pd_7WTe_2$ is locked to the crystal direction of the single-crystalline $WTe_2$ monolayer seed.

### *Generalization of the mass transport*

We next demonstrate that this phenomenon of 2D crystal growth within a nanoconfined 2D space on an atomically thin crystal is not unique to Pd on $WTe_2$; instead, it is highly generalizable. In Extended Data Fig. 3 and Extended Data Figs. 9-11, we demonstrate that similar Pd transport occurs for bilayer $WTe_2$ (see also Methods). In short, we find that the thickness of the synthesized material from a bilayer $WTe_2$ is twice that of a monolayer, with the same atomic composition (i.e., $Pd_7WTe_2$). In some regions of a bilayer-seeded sample, STEM reveals a hexagonal lattice consistent with monolayer $Pd_7WTe_2$. Yet, we also found different structures and domains formed on bilayer, including regions where we see moiré patterns (Extended Data Fig. 11). This is distinct from monolayer case, where we find highly uniform structure over a very large area. Future experiments are required to fully characterize the bilayer-seeded case, which may yield interesting possibilities, including the potential creation of new moiré materials.

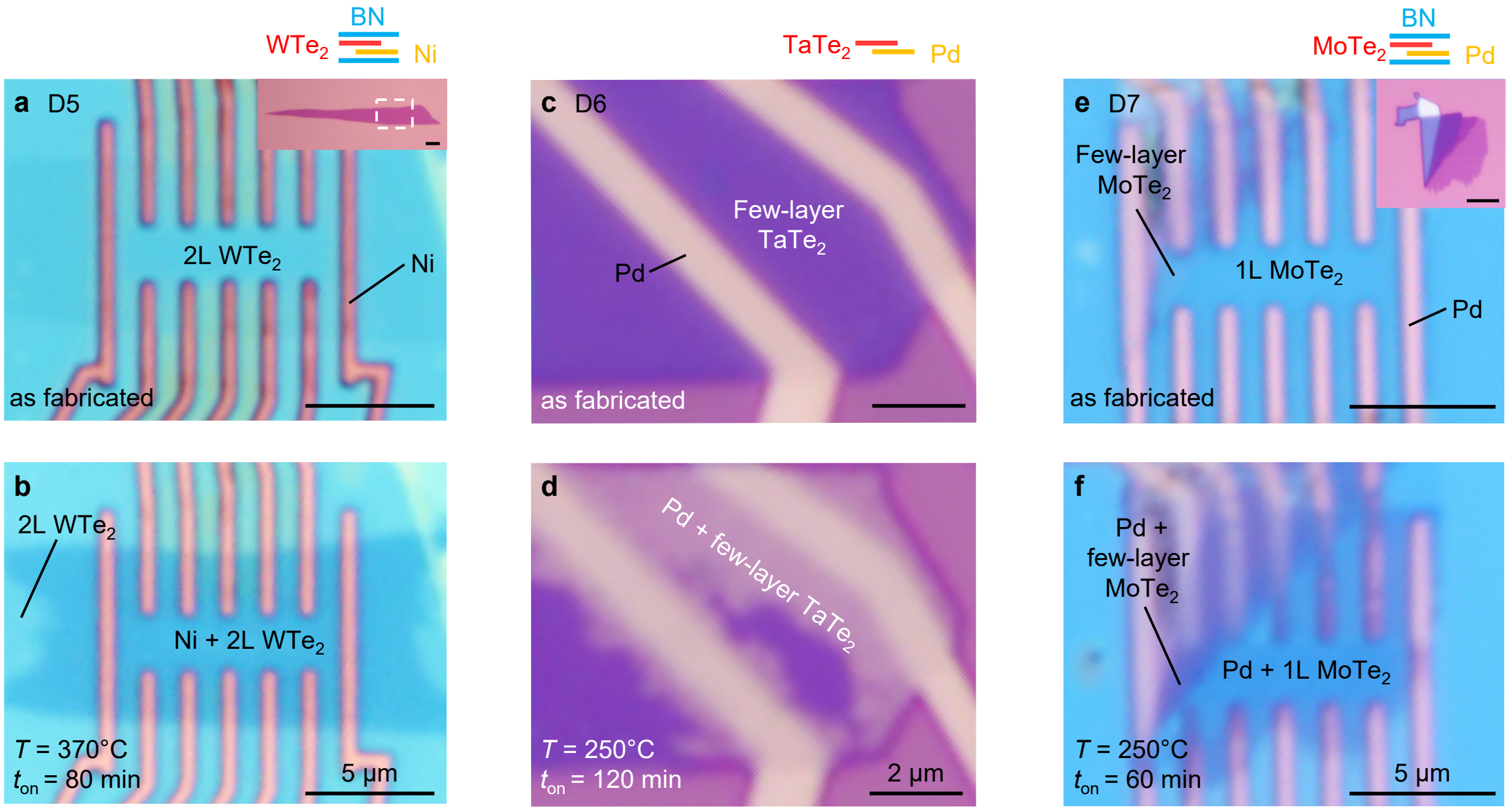

**Figure 4 | Generalization of the unusual mass transport. a,** Optical image of device D5 with bilayer $WTe_2$ and Ni sources encapsulated by hBN, prior to thermal treatment. Insets on top show an image of bilayer $WTe_2$ as exfoliated, as well as the stack structure. **b,** Optical image of the same device after being held at ~ 370 °C for 80 minutes. Ni transport was induced on and only on $WTe_2$, visualized as the optical darker region. The critical temperature to turn on Ni transport on bilayer $WTe_2$ is 350 °C and the stack is held at slightly higher temperature (370 °C) to speed up the process. **c,** Optical image of D6 with few-layer $TaTe_2$ and Pd contacts, as fabricated. The inset displays the stack structure. **d,** Optical image of the same device after being held at 250 °C for 120 minutes. The color of $TaTe_2$ changed from purple to yellow, which indicates the transport of Pd and formation of new materials. **e,** Optical image of device D7 with both monolayer and few-layer $MoTe_2$ as well as Pd contacts. The as-exfoliated $MoTe_2$ flake and the stack structure are shown as the inset. **f,** Optical image of the same device after being held at 250 °C for 60 minutes. The color of both monolayer and few-layer $MoTe_2$ become darker due to Pd transport.

Figure 4 illustrates three additional examples. The first is the observation of a similar long-distance transport of nickel (Ni) on bilayer $WTe_2$, at a temperature of ~ 370 ºC, again well below the known melting point of Ni (1,455 ºC). This can be clearly visualized by the color change after the thermal treatment in the optical images of the vdW stack under a microscope, as shown in Figs. 4a & b. The second is the similar transport of Pd on a few-layer $TaTe_2$ at a temperature of ~ 250 ºC (Figs. 4c & d). Here the optical contrast of the resulting material is brighter than that of monolayer- or bilayer-seeded samples since we used a few-layer flake, indicating that more Pd transports into the few layer samples. In the third example, we show that Pd transports similarly along a monolayer $MoTe_2$ (Figs. 4e & f), where Pd resides on the entire flake after the vdW stack is being held at 250 ºC for 60 minutes. It seems that all these observations involve melting point depression[29] of nanoscale metals, reactive 2D interfaces and chemical interactions. The non-Fickian transport characteristics (Fig. 1) imply that simple diffusion cannot explain the situation and we believe that chemical affinity plays a key role. Future efforts are necessary to reveal the exact mechanism of such anomalous mass transport in various cases observed here. The understanding may provide new insights on nanoscale atomic transport. If the transporting metal is indeed partially in a liquid phase, our experiment then realizes an atomically thin 2D liquid metal, a highly intriguing situation whose realization was not known previously. Overall, our experimental results imply that the anomalous non-Fickian transport phenomenon is, surprisingly, quite ubiquitous for various situations of metals on existing 2D crystals.

## Summary

Our work demonstrates a route for controlled chemistry on a monolayer. We envision that the nanoconfined growth mechanism templated on monolayer crystals will largely expand the library of 2D materials[1,2,9] and their functionalities, especially since it provides a new strategy for synthesizing 2D materials inaccessible using conventional approaches. In contrast to conventional 2D growth techniques, the approach here is highly compatible with existing mature techniques for creating high quality 2D devices, including vdW encapsulation and nanodevice fabrication, a big advantage for further investigating and engineering their physical properties and device functionalities. The exploration along these directions may enable advances in the study of 2D superconductors, magnets, and topological states of matter.

## Acknowledgments

This work is mainly supported by ONR through a Young Investigator Award (N00014-21-1-2804) to S.W. STEM measurements and device fabrications are supported by NSF through a CAREER award (DMR-1942942) to S.W. and the Materials Research Science and Engineering Center (MRSEC) program of the National Science Foundation (DMR-2011750) through support to L. M. S. and S.W. Device fabrication is partially supported by AFOSR Young Investigator program (FA9550-23-1-0140) to S.W. S.W. and L.M.S. acknowledge support from the Eric and Wendy Schmidt Transformative Technology Fund at Princeton. L.M.S. acknowledges support from the Gordon and Betty Moore Foundation through Grants GBMF9064 and the David and Lucile Packard Foundation and the Sloan Foundation. Y.J. acknowledges support from the Princeton Charlotte Elizabeth Procter Fellowship program. T.S. acknowledges support from the Princeton Physics Dicke Fellowship program. A.J.U acknowledges support from the Rothschild Foundation and the Zuckerman Foundation. The authors acknowledge the use of Princeton's Imaging and Analysis Center, which is partially supported by the Princeton Center for Complex Materials, a National Science Foundation NSF-MRSEC program (DMR-2011750). K.W. and T.T. acknowledge support from the JSPS KAKENHI (Grant Numbers 20H00354, 21H05233 and 23H02052) and World Premier International Research Center Initiative (WPI), MEXT, Japan.

## Author contributions

Y. J. and S.W. co-discovered the phenomena. Y.J. fabricated all the devices, performed optical and transport measurements, and analyzed data, assisted by Y.T., G.Y., T.S., P.W., A.J.U and M.O. and supervised by S. W. Y.J. and F. Y. co-developed the STEM fabrication and measurement procedures, supervised by L.S. and S.W., in collaboration with G.C. and N.Y. R.S. and L.M.S. grew and characterized bulk $WTe_2$ crystals. K.W. and T.T. provided hBN crystals. S.W. and Y. J. interpreted the results and wrote the paper with input from all authors.

## Competing interests

Authors declare that they have no competing interests.

## Data availability

All data needed to evaluate the conclusions in the paper are present in the paper. Additional data related to this paper are available from the corresponding author upon reasonable request.

## Methods

### Device Fabrication

vdW stacks

The high-quality $WTe_2$ bulk crystals were grown using the methods described in previous works[28,31]. hBN flakes were exfoliated on $SiO_2$/Si substrates and identified under an optical microscope. hBN flakes of appropriate thickness (10 – 20 nm) were characterized using AFM before being transferred onto undoped $SiO_2$/Si wafers with pre-patterned metal alignment markers, using standard dry transfer techniques. We then employed electron beam lithography, followed by cold development, reactive ion etching, and metal deposition, to deposit the patterned Pd (Ni for D5) sources (typically 10 ~ 20 nm thick; occasionally with additional ~ 3 nm Ti as a sticking layer) on the hBN flakes (except for D6, whose metal source is directly deposited on $SiO_2$/Si). Such prepared bottom stacks were then cleaned with an AFM tip. The $WTe_2$ monolayers were exfoliated and transferred onto the prepared bottoms. The details of the procedure can be found in our previous work[27,28]. The final vdW stacks were then placed on a hot plate for Pd transport, while being monitored under an optical microscope. The whole process involving $WTe_2$ was performed in glovebox with $H_2O$ < 0.1 ppm and $O_2$ < 0.1 ppm. Similar process was used to create the devices of different seeds ($TaTe_2$ & $MoTe_2$) and metal (Ni). .

TEM cross-section devices

The creation of TEM cross-section devices started with the vdW stacks produced using the above-mentioned method. After Pd transport, the stacks were then coated with 50 nm thick amorphous carbon by Sputter Coater as a protective layer before being loaded into Helios NanoLab G3 UC dual-beam focused ion beam and scanning electron microscope (FIB/SEM) system. After that, a standard lift-out process was performed within the FIB system to cut out a lamella specimen from the stack, transfer the specimen onto a TEM grid, and polish its cross-section until it is electron transparent (thickness ~ 50 nm). Sample thinning was accomplished by gently polishing the sample using a 2 kV $Ga^+$ ion beam in order to minimize surface damage caused by the ion beam. Finally, the specimen is quickly transferred into TEM for high-resolution imaging.

Suspended TEM devices

Standard TEM grids with holey silicon nitride supporting film were first treated with oxygen plasma for 5 minutes on both sides and then deposited with 20 nm Pd. After that, the $WTe_2$ monolayers were exfoliated and transferred onto the TEM grids with standard dry transfer techniques. The grids, covered with $WTe_2$ and the polycarbonate (PC) used for the transfer process, were then placed on a hot plate with the temperature maintained at ~ 190 °C for 5mins. The PC was removed by immersing the grids in chloroform for 1 hour in glovebox afterwards. The TEM grids were then mounted onto a Gatan double tilt vacuum transfer holder. The exfoliation, stacking, crystal growth, PC removal, and sample mounting are all performed within the glovebox with $H_2O$ < 0.1 ppm and $O_2$ < 0.1 ppm. The devices were also well protected from air by the vacuum transfer holder (filled with Ar) when it is transferred from the glovebox to the TEM instrument.

### Optical and AFM Measurements

The optical measurements were carried out with a home-built transfer setup that is equipped with a heating table (hot plate), Nikon Eclipse LV150N microscope and Canon EOS 5D Mark IV DSLR camera. The video shown in Supplementary Video 1 was recorded by software OBS Studio and edited with VideoProc Vlogger. Optical images displayed in this work were either directly captured by the camera or extracted from the recorded video. The AFM data in this work were obtained from either Bruker Dimension Edge or Bruker Dimension Icon and analyzed with software NanoScope Analysis.

**Microstructure Characterization**

Atomic resolution high-angle annular dark-field (HAADF) STEM imaging and energy dispersive X-ray spectroscopy (EDX) mapping were performed on a Titan Cubed Themis 300 double Cs-corrected scanning/transmission electron microscope (S/TEM), equipped with an extreme field emission gun source and a super-X EDS system. The system was operated at 300 kV.

**Summary of device parameters (Table 1)**

| | D1 | D2 | D3 | D4 | D5 | D6 | D7 |
|---|---|---|---|---|---|---|---|
| Top hBN | 7 nm | 10 nm | 23 nm | | ~ 15 nm | | ~ 13 nm |
| Precursor | 1L $WTe_2$ | 1L & 2L $WTe_2$ | 1L & few-layer $WTe_2$ | 1L $WTe_2$ | 2L $WTe_2$ | Few-layer $TaTe_2$ | 1L & few-layer $MoTe_2$ |
| Metal source | Pd | Pd | Pd | Pd | Ni | Pd | Pd |
| Bottom hBN | 23 nm | 9 nm | 22 nm | | 12 nm | | 18 nm |

**Atomic modeling of the new 2D material $Pd_7WTe_2$**

Since the material is Pd rich, we first compare the STEM observation with the lattice of crystalline, elemental Pd. The face-centered cubic (FCC) structure of Pd displays a hexagonal lattice when viewed along both the [111] and [110] directions. (*i*) Along the [111] direction, equilateral hexagonal lattices are closed-packed with A-B-C-A stacking order (Extended Data Figs. 6a & b). In the plot, Pd atoms in different layers are colored differently to illustrate the stacking order. One finds that the interatomic spacing is significantly smaller than that of $Pd_7WTe_2$ (0.275 nm < 0.44 nm). Indeed, we simulated the diffraction pattern of atomically thin Pd viewed along the [111] direction, which is clearly different from the experimentally observed version of $Pd_7WTe_2$ (Fig. 3d). (*ii*) If Pd was viewed along the [110] direction, the structure consists of stacked rectangular lattices with an A–B–A order, resulting in a distorted hexagonal lattice (Extended Data Figs. 6c & d). The distortion of hexagon is further revealed in the simulated diffraction pattern, which shows different periodicities (0.19 nm and 0.22 nm) along two high-symmetry directions. While the observed interatomic distance in this arrangement is much closer to the measured value in $Pd_7WTe_2$ (0.389 nm vs 0.38 nm), the symmetry of the pattern does not satisfactorily describe that of $Pd_7WTe_2$, which maintains the six-fold symmetry (Fig. 3d). It seems thus that the Pd lattice observed within $Pd_7WTe_2$, in the plan-view, resembles a significantly strained Pd [111] lattice (equivalent to ~ 60% strain!). This is dramatically larger as in other reported strained Pd nanocrystals, such as the Pd islands found on $PdCoO_2$ (~ 4 % strain)[32].

In Extended Data Figs. 6e & f, we propose a model for describing the observed $Pd_7WTe_2$ lattice structure. The plan-view and cross-section of a proposed crystal structure for $Pd_7WTe_2$ are shown respectively. The material consists of 10 atomic layers, each with a hexagonal lattice with a lattice constant of 0.44 nm. In the vertical direction, the 10 atomic layers are close-packed. In the model, we assume that the top three layers are Te-W-Te, arranged also in the hexagonal lattice (like the 1T monolayer structure), with a stacking order of A-B-C. The bottom 7 layers are occupied by Pd atoms and stacked in an A-X-A (X can be either B or C with equal probability) order, as illustrated in Extended Data Figs. 6e & f. All characteristic lengths, in both top and side views, are consistent with the observations in Figs. 2 and 3. The simulated diffraction pattern also agrees with our experimental observation in Fig. 3d. More details of the model may be resolved with future experiments.

**Pd transport on bilayer $WTe_2$**

*Pd transport process* **-** The transport of Pd on bilayer $WTe_2$ is investigated using device D2, which contains both monolayer and bilayer $WTe_2$ regions as well as separated Pd sources. The transportation process is illustrated through optical images in Extended Data Fig. 9. Two noticeable effects were observed during the process: (1) Pd spreads on the bilayer at a lower critical temperature, ~ 170 °C, compared to the monolayer ~ 190 °C (Extended Data Fig. 9c). (2) An interesting "diode effect" is observed in the propagation of Pd across monolayer/bilayer step. Namely, Pd can freely transport from monolayer to bilayer but not vice versa, as illustrated in Extended Data Fig. 9d. We currently don't have a comprehensive understanding of these interesting observations.

*Bilayer-seeded $Pd_7WTe_2$* **-** In Extended Data Fig. 3d, the thickness of the resulting materials from Pd transport on bilayer $WTe_2$ is measured to be ~ 3.1 nm in AFM, which is double the thickness of monolayer $Pd_7WTe_2$ (~ 1.5 nm in AFM). This double thickness relation is further confirmed by STEM measurement in Extended Data Fig. 10, where HAADF STEM cross-section images of bilayer $WTe_2$ (S1) and the resulting material (S2) are shown. EDX analysis on both cross-section (Extended Data Fig. 10d) and plan-view (Extended Data Fig. 11e) confirm the same atomic ratio of Pd:W:Te, close to 7:1:2, the same as the monolayer-seeded $Pd_7WTe_2$. The in-plane lattice structure of the bilayer-seeded $Pd_7WTe_2$ is examined in Extended Data Fig. 11. In some regions, we find a crystalline structure consistent with the hexagonal lattice of the monolayer-seeded $Pd_7WTe_2$ (Extended Data Fig. 11a), whereas we also find regions that display various other patterns, including moiré patterns (Extended Data Fig. 11b-d). This inhomogeneity in bilayer-seeded suspended sample may be caused during the fabrication process. In monolayer-seeded samples, we typically found uniform crystalline structure. Further experiments are necessary to fully understand the situation of bilayer-seeded growth.

# Extended Data Figures

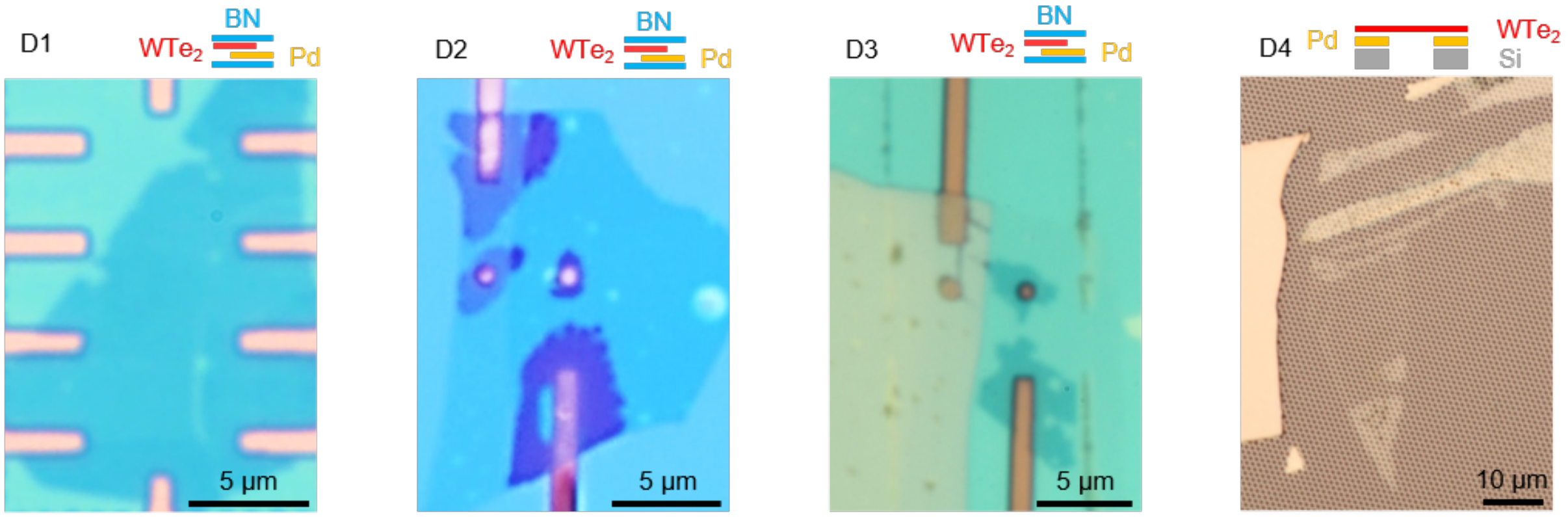

**Extended Data Fig. 1 | Devices images.** Optical images of $Pd_7WTe_2$ devices D1-D4 used in this study are summarized here. Schematic diagrams of the structure of each device (before growth) are sketched next to the image.

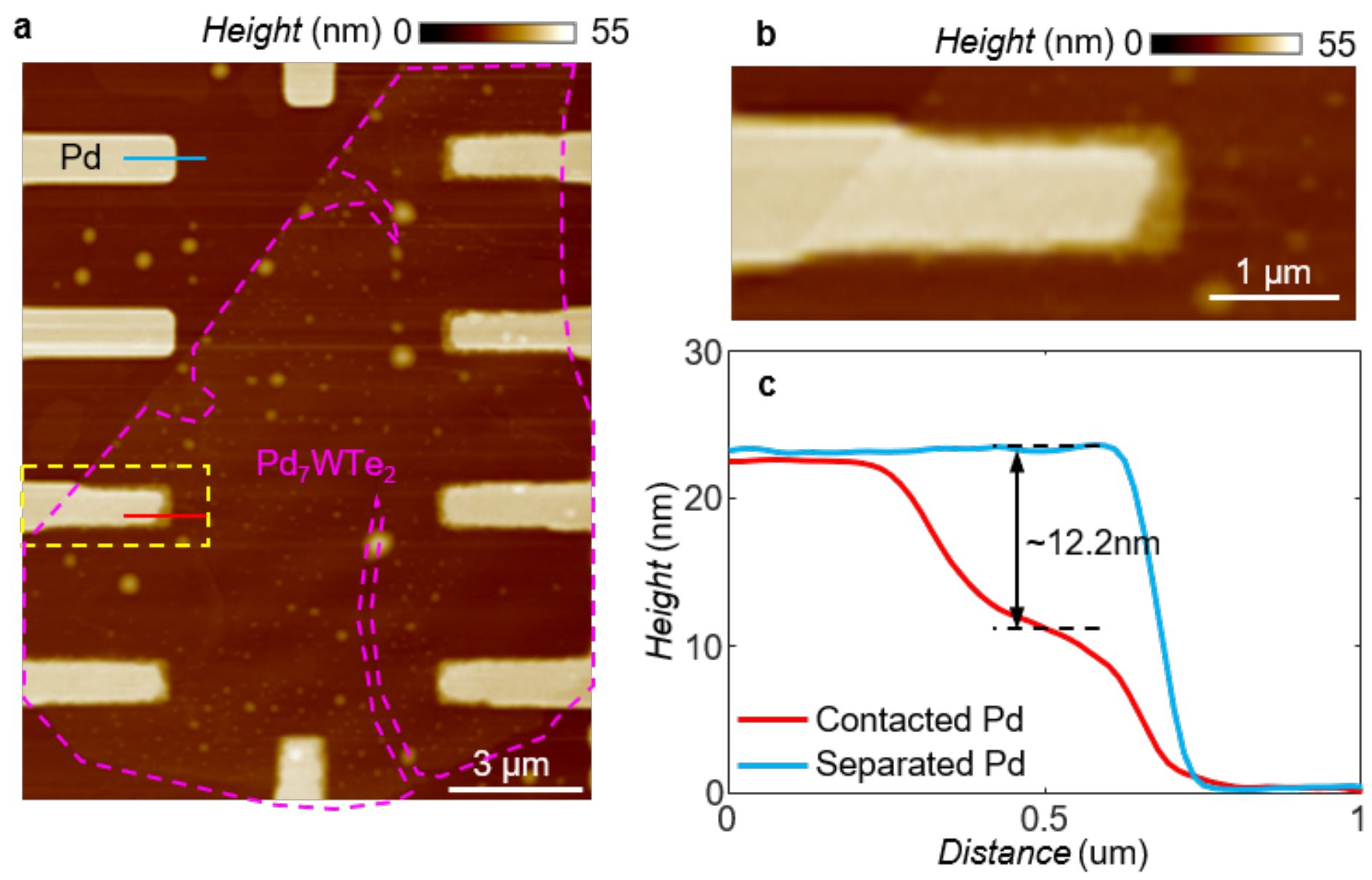

**Extended Data Fig. 2 | AFM characterization of D1 after Pd transport. a,** The AFM height image, with $Pd_7WTe_2$ crystal outlined in purple. The Pd islands covered by $WTe_2$ become narrower and thinner, while no visible change is seen for the separated Pd islands. **b,** A zoomed-in image of a Pd island, highlighted by the yellow dash line in **a**. **c,** Height profiles of the Pd islands along the indicated lines in **a**.

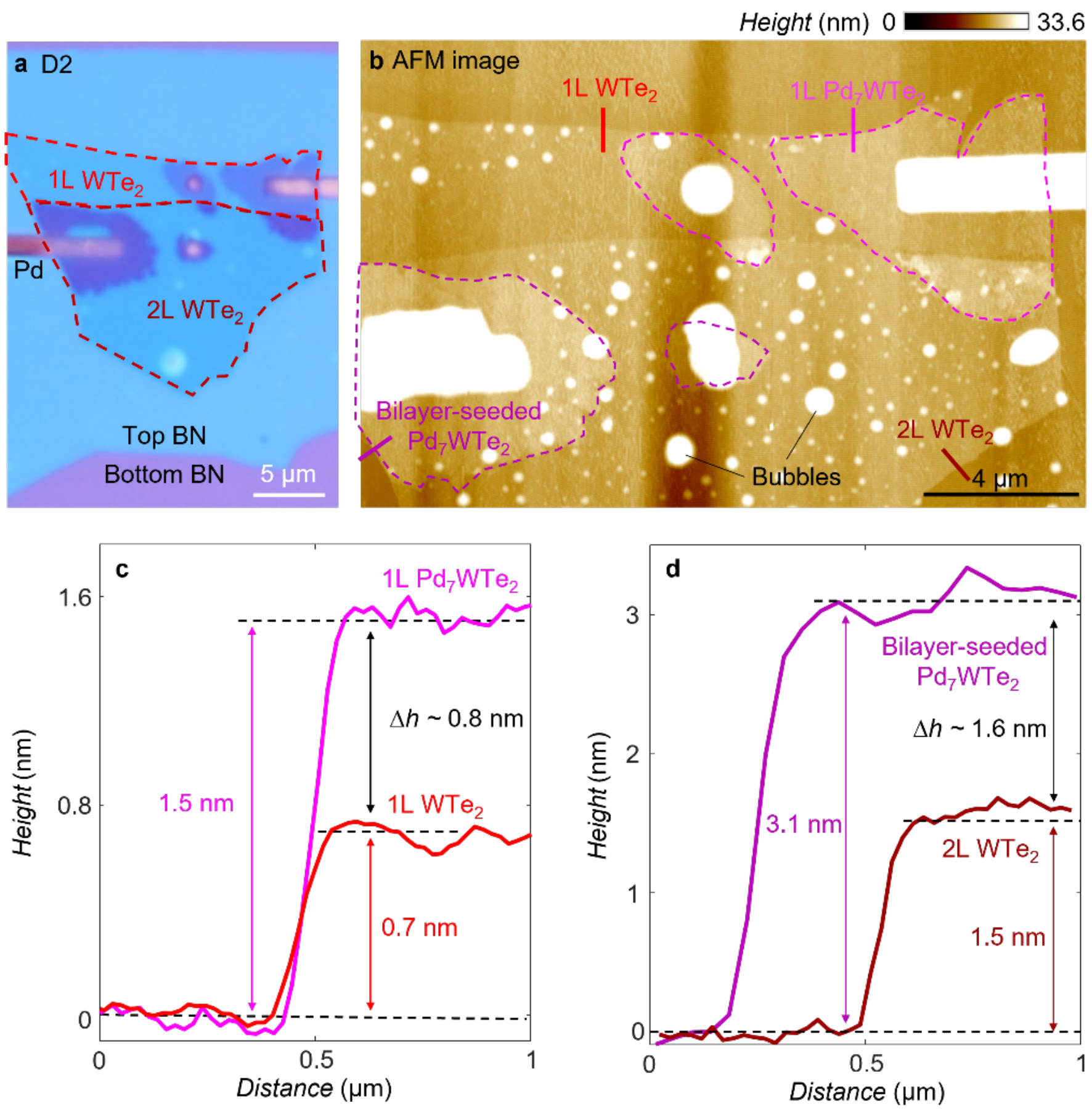

**Extended Data Fig. 3 | Optical and AFM characterizations of D2. a,** An optical image of D2 after Pd transport. **b,** An AFM height image of D2 with $Pd_7WTe_2$ outlined in purple dash lines. The locations of height profiles measured in **c & d** are indicated by solid lines. **c,** The height profile of monolayer $WTe_2$ and $Pd_7WTe_2$. **d,** The height profile of bilayer $WTe_2$ and bilayer-seeded $Pd_7WTe_2$. The height of the bilayer-seeded $Pd_7WTe_2$ is twice that of the monolayer-seeded $Pd_7WTe_2$.

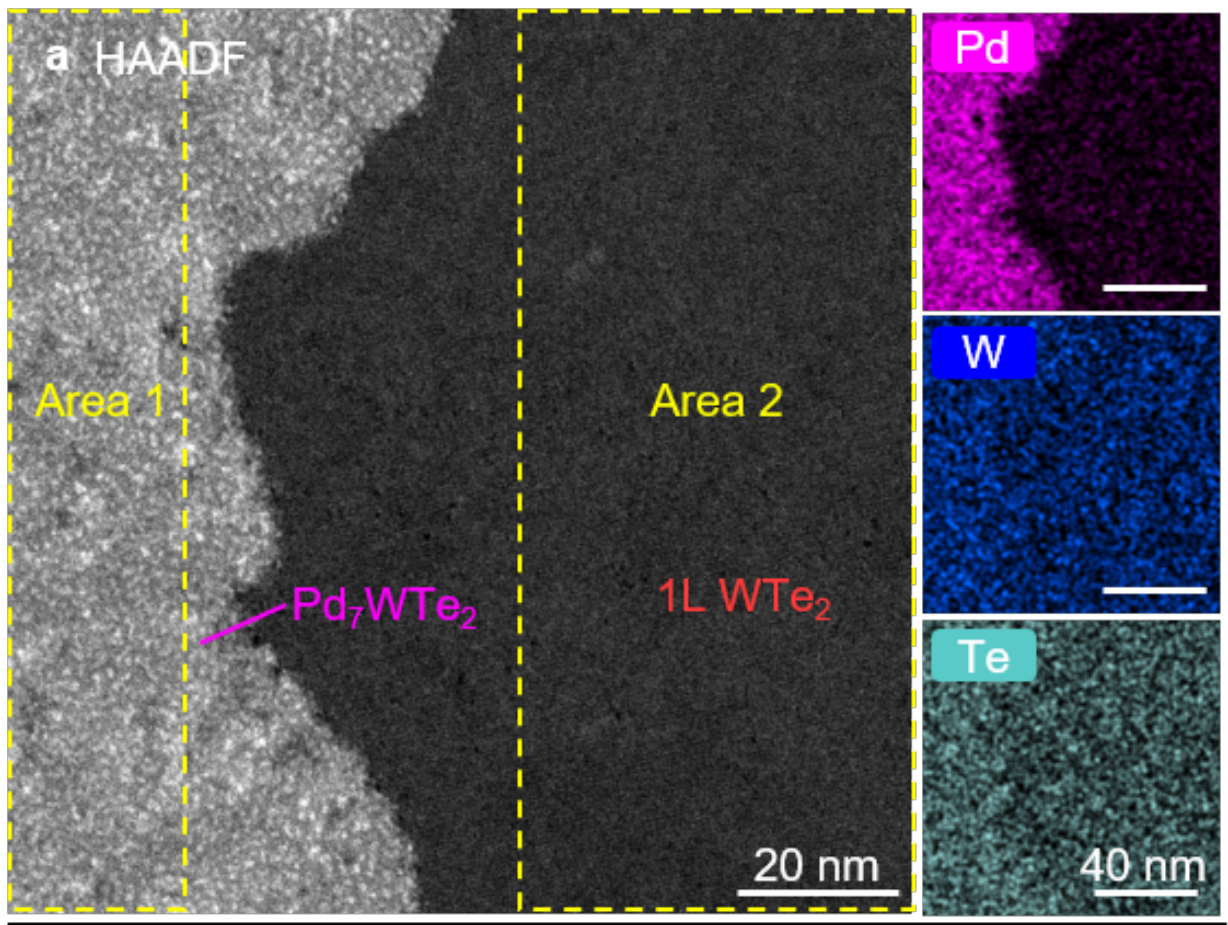

| D4 Spot2 | Atomic fraction (%) | | |
|---|---|---|---|
| | Pd | W | Te |
| Area 1 | 68.8 | 11.1 | 20.1 |
| Area 2 | 1.6 | 32.6 | 65.8 |

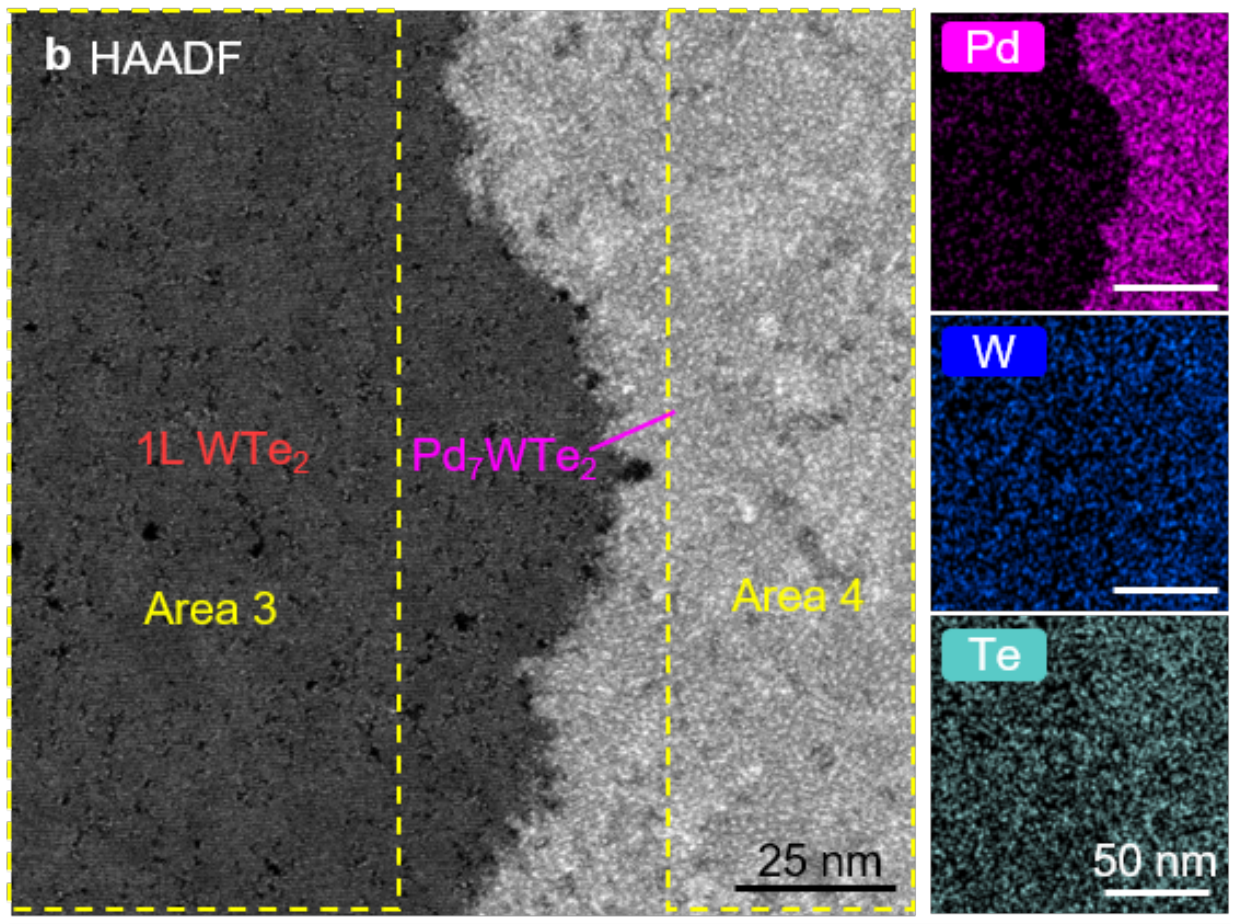

| D4 Spot3 | Atomic fraction (%) | | |
|---|---|---|---|
| | Pd | W | Te |
| Area 3 | 2.0 | 33.0 | 65.0 |
| Area 4 | 68.8 | 10.3 | 20.9 |

**Extended Data Fig. 4 | Additional EDX analyses and elemental mappings on D4. a & b,** The HAADF STEM images and corresponding EDX elemental mappings on monolayer $WTe_2$ and $Pd_7WTe_2$ at two additional locations on D4. The extracted atomic fractions of selected areas are summarized in the tables. The Pd:W:Te atomic ratios in Pd propagated regions are always found to be approximately 7:1:2.

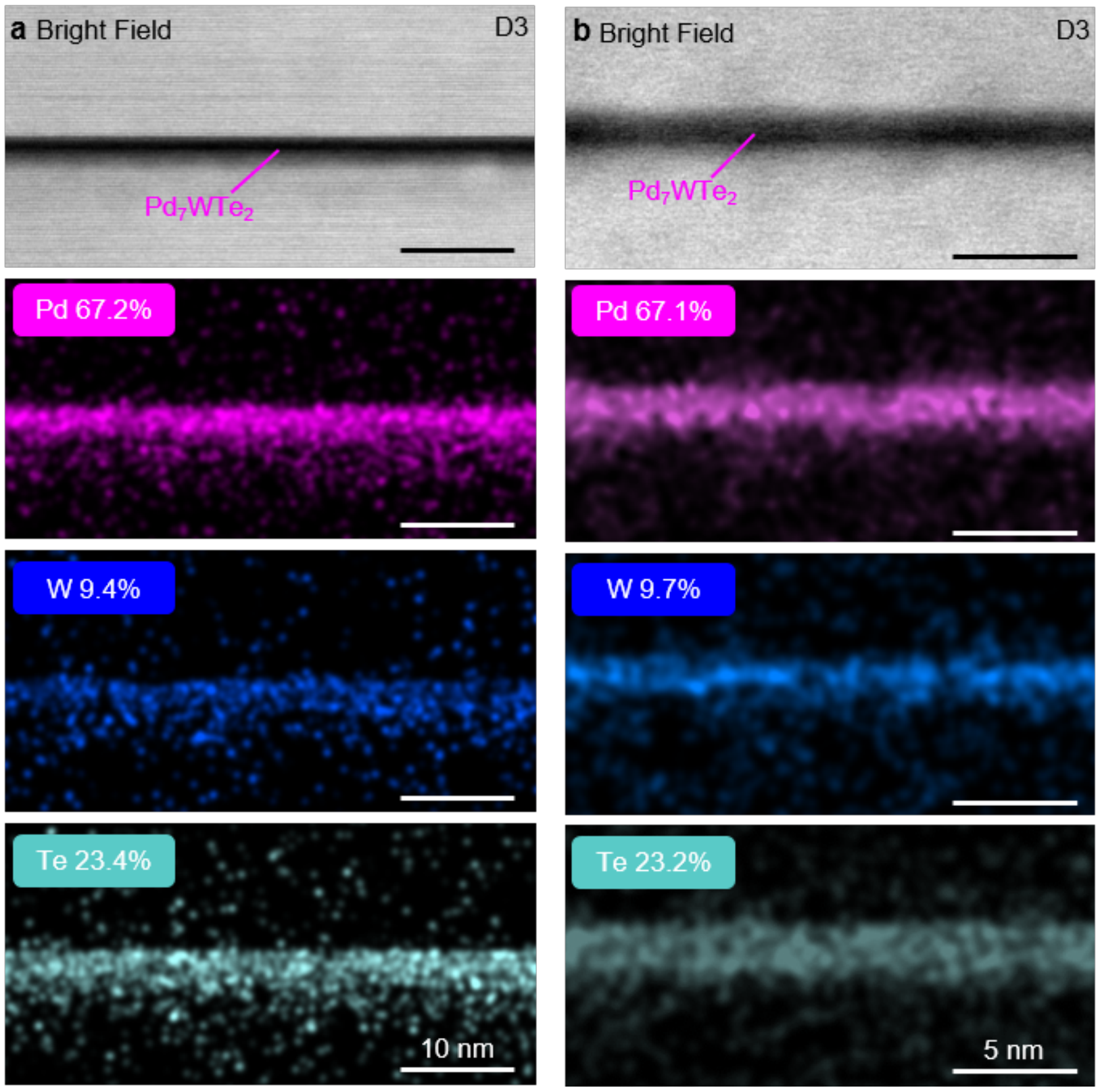

**Extended Data Fig. 5 | Additional EDX analyses and elemental mappings on the cross-sections of D3. a & b,** The bright field images and corresponding EDX elemental mappings on another 2 locations of D3 with $Pd_7WTe_2$ synthesized from monolayer $WTe_2$. The Pd:W:Te atomic ratio is found to be close to 7:1:2. The small deviations may be the result of insufficient signal intensity in the cross-section measurements.

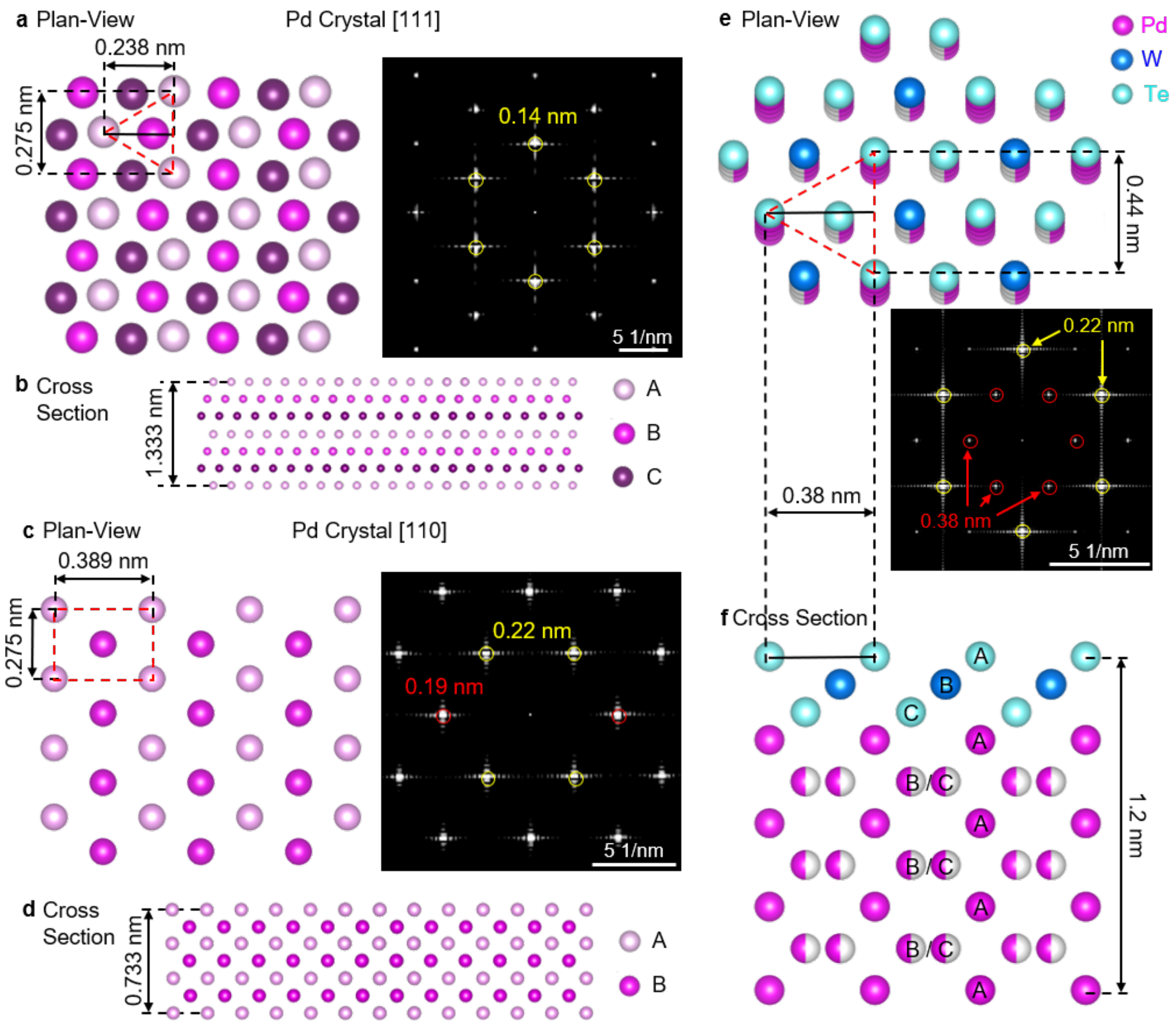

**Extended Data Fig. 6 | Bulk Pd crystal *v.s.* the proposed crystal structure of $Pd_7WTe_2$. a,** An FCC bulk Pd crystal viewed along the [111] direction. Pd atoms from different layers are colored in different shades. The plan-view of such structure exhibits six-fold symmetry, as is shown by the simulated diffraction pattern (right panel). **b,** Cross-section of 7-layers of Pd atoms organized in the structure corresponding to **a**. **c,** The Pd crystal viewed along the [110] direction, which consists of alternatively stacked (A-B-A) rectangular lattices. Again, Pd atoms from different layers are shaded differently for a better visualization. The resulting plan-view lacks six-fold symmetry. The distorted hexagon is clearly seen in the simulated diffraction pattern (right panel). **d,** Cross-section of 7-layers of Pd atoms organized in the structure corresponding to **c**. **e & f,** The plan-view (top panel) and cross-section (bottom panel) of our proposed crystal structure for $Pd_7WTe_2$, consisting of 10 layers of hexagonal lattices, with the stacking order and atom species indicated. The half-colored atoms indicate that these sites are only filled 50 %. The simulated plan-view diffraction pattern (middle panel) captures characteristic features of experimental results.

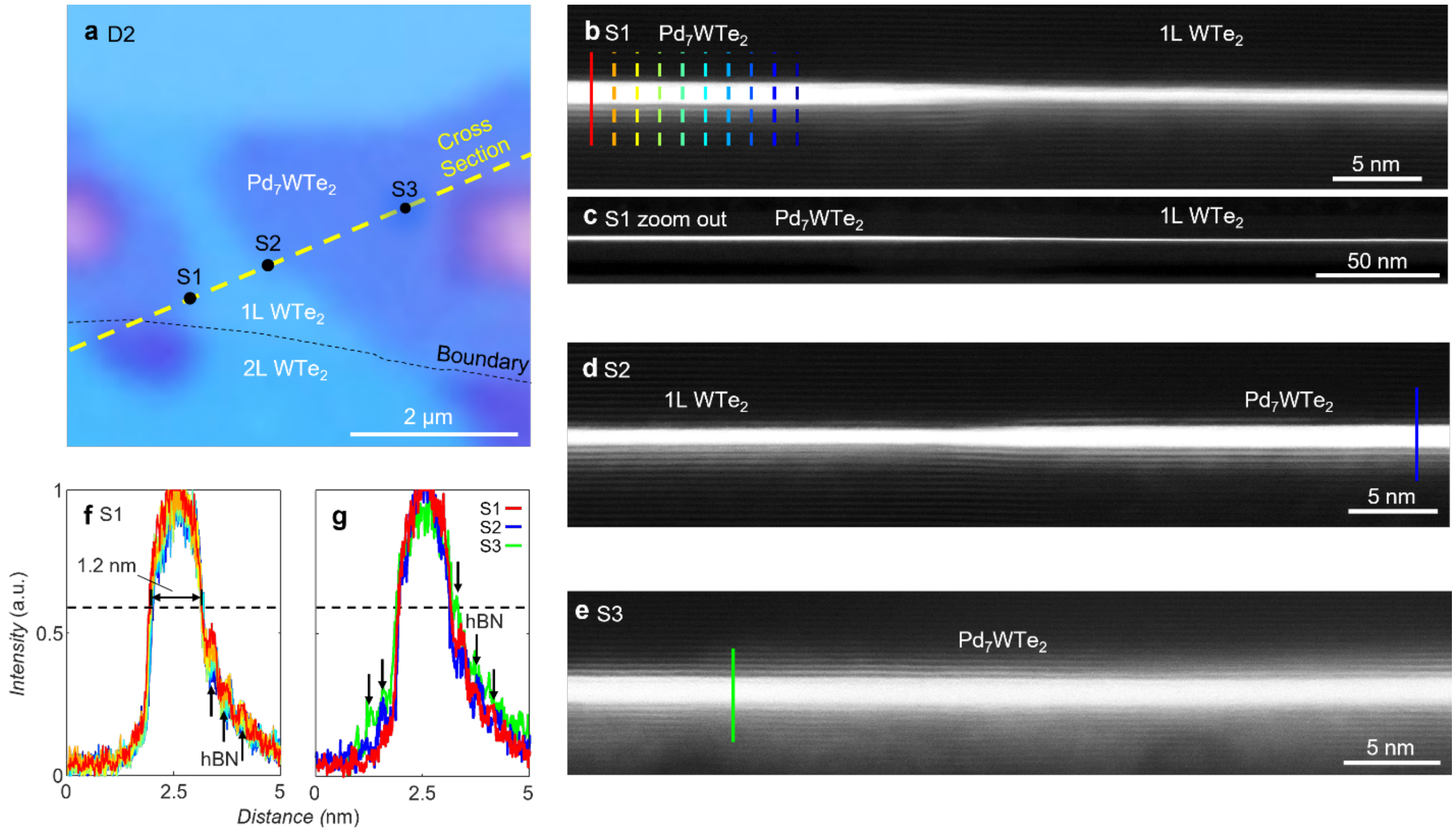

**Extended Data Fig. 7 | Uniform thickness of monolayer seeded $Pd_7WTe_2$. a,** An optical image of the target region of device D2, focusing on the monolayer $WTe_2$ and $Pd_7WTe_2$ areas. STEM measurements were performed along the cross-section cut indicated by the yellow dashed line. Spots S1, S2 and S3 are estimated locations under examination in **b**-**e**. **b,** A HAADF STEM image captured at spot S1, which interfaces monolayer $WTe_2$ and $Pd_7WTe_2$. The extracted intensity along the colored line-cuts is shown in **f**. **c,** The zoomed-out STEM image at spot S1, showing the uniform thickness on the scale of ~ 100 nm. **d,** A HAADF STEM image acquired at the interface between monolayer $Pd_7WTe_2$ and $WTe_2$ at spot S2. **e,** A HAADF STEM image obtained at spot S3. **f,** The extracted intensity from S1 along the line-cuts in **b**, indicated by the vertical lines. The signal from hBN layer structure is indicated by arrows. **g,** The extracted intensity from S1-S3 along line-cuts indicated by the solid lines in **b**, **d** & **e**. The data confirms that the thickness of $Pd_7WTe_2$ is uniform on the same monolayer seed, even the Pd comes from separate sources.

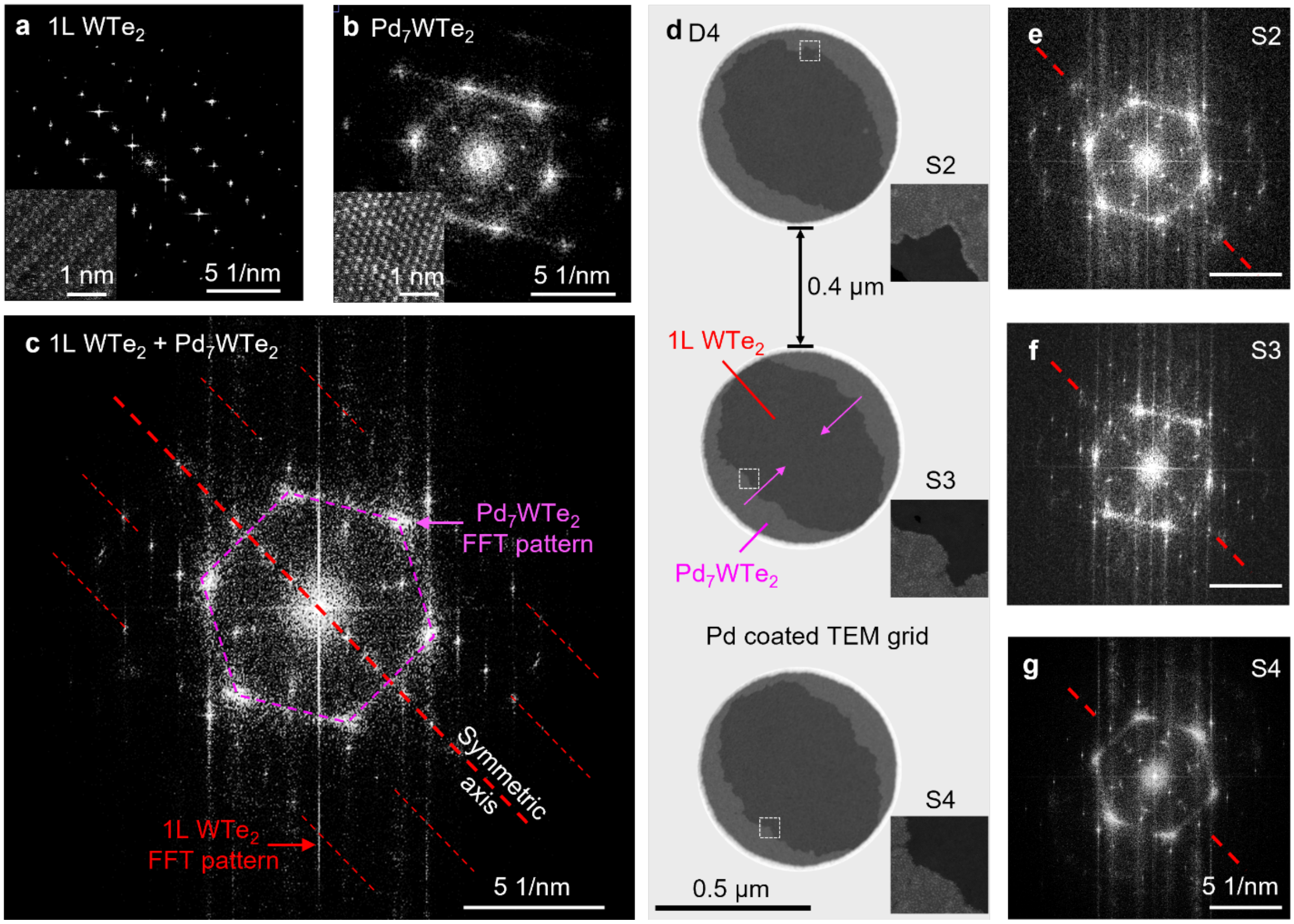

**Extended Data Fig. 8 | The locked orientation of $Pd_7WTe_2$ lattices and monolayer $WTe_2$. a,** The FFT pattern and atomic-resolution image (inset) of monolayer $WTe_2$ crystal. **b,** The FFT pattern and atomic-resolution image (inset) of monolayer seeded $Pd_7WTe_2$. **c,** An FFT pattern of the interfacing region (S1, Fig. 3b) including both monolayer $WTe_2$ and $Pd_7WTe_2$. The resulting FFT pattern exhibits characteristic features of both materials, highlighted separately by dashed lines (red for $WTe_2$, purple for $Pd_7WTe_2$). It is noted that their FFT patterns are aligned with specific angles and symmetric about the indicated symmetric axis. **d,** STEM images of another three additional spots, S2-S4, selected from three separated TEM grid holes. The locations of the three spots are indicated by white dashed boxes. **e – g,** The FFT patterns of S2-S4, which exhibit same orientation relation between monolayer $WTe_2$ and $Pd_7WTe_2$ lattice. The data shows that the $Pd_7WTe_2$ lattice orientation is determined by the seed monolayer $WTe_2$.

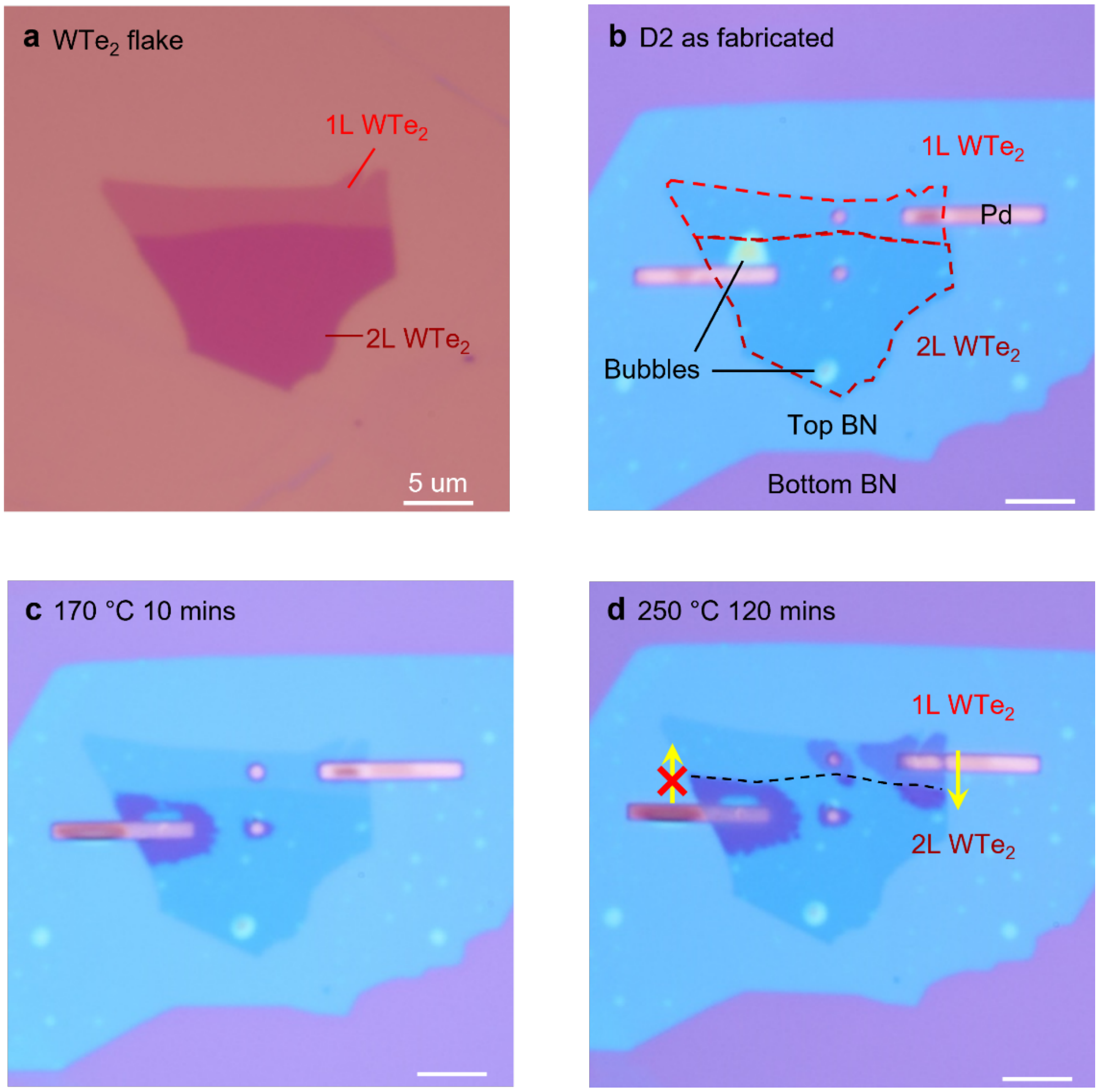

**Extended Data Fig. 9 | Pd transport on monolayer and bilayer $WTe_2$. a,** Optical image of a $WTe_2$ flake with both monolayer and bilayer regions. **b,** Optical image of device D2 as fabricated, where monolayer and bilayer $WTe_2$ are in contact with separate Pd sources. **c,** Optical image of D2 after being held at 170 °C for 10 minutes, where Pd spreads significantly on bilayer $WTe_2$ while not noticeable on monolayer $WTe_2$. **d,** Optical image of D2 after being held at 250 °C for 120 minutes, showing Pd transport from monolayer $WTe_2$ to bilayer $WTe_2$, but not vice versa.

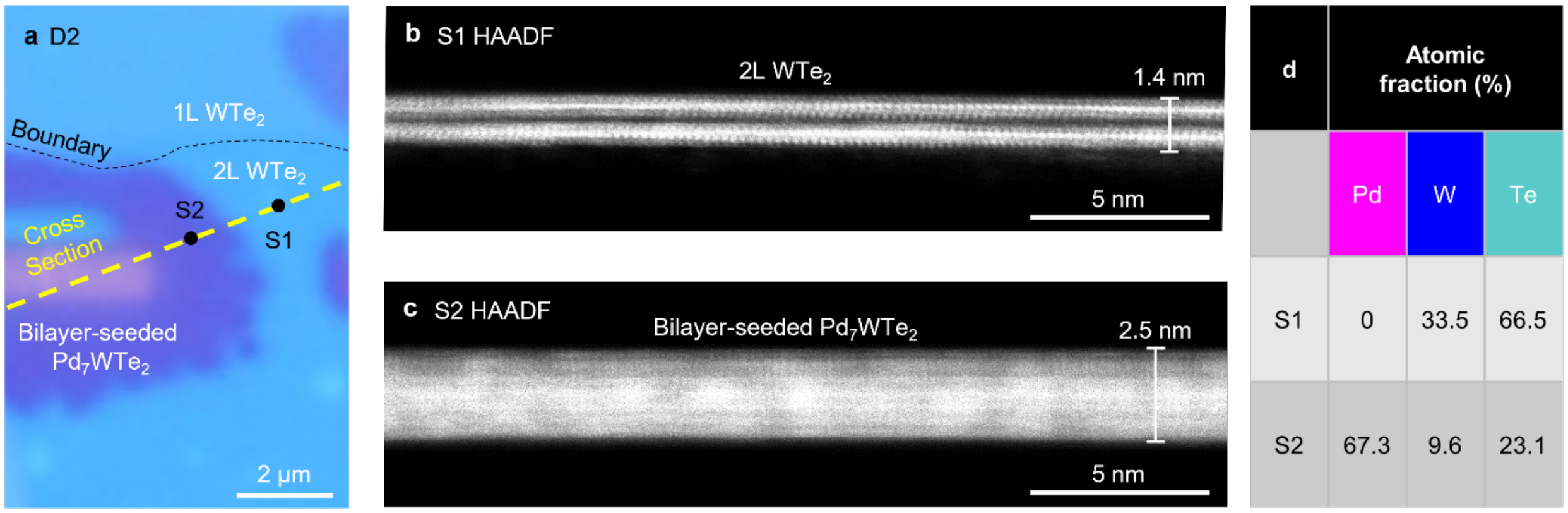

**Extended Data Fig. 10 | STEM characterization of a bilayer-seeded** $Pd_7WTe_2$. **a,** A zoom-in optical image of device D2, focusing on the bilayer $WTe_2$ and bilayer-seeded $Pd_7WTe_2$ areas. STEM measurements were performed along the cross-section line (yellow line) and images were obtained at estimated spots S1 and S2, as indicated. **b,** HAADF STEM cross-section image of bilayer $WTe_2$ at location S1, where the vdW gap is clearly observed. **c,** HAADF STEM cross-section image of bilayer-seeded $Pd_7WTe_2$ at location S2. The thickness is measured to be 2.5 nm, double the thickness of monolayer-seeded $Pd_7WTe_2$ (~ 1.2 nm). **d,** Atomic fractions for bilayer $WTe_2$ and bilayer-seeded $Pd_7WTe_2$ from EDX analysis. We did not obtain atomic resolution in the STEM images of the bilayer-seeded $Pd_7WTe_2$ region, which we suspect might be a result of in-plane domain inhomogeneity. Future improvements are necessary for obtaining an image with atomic resolution.

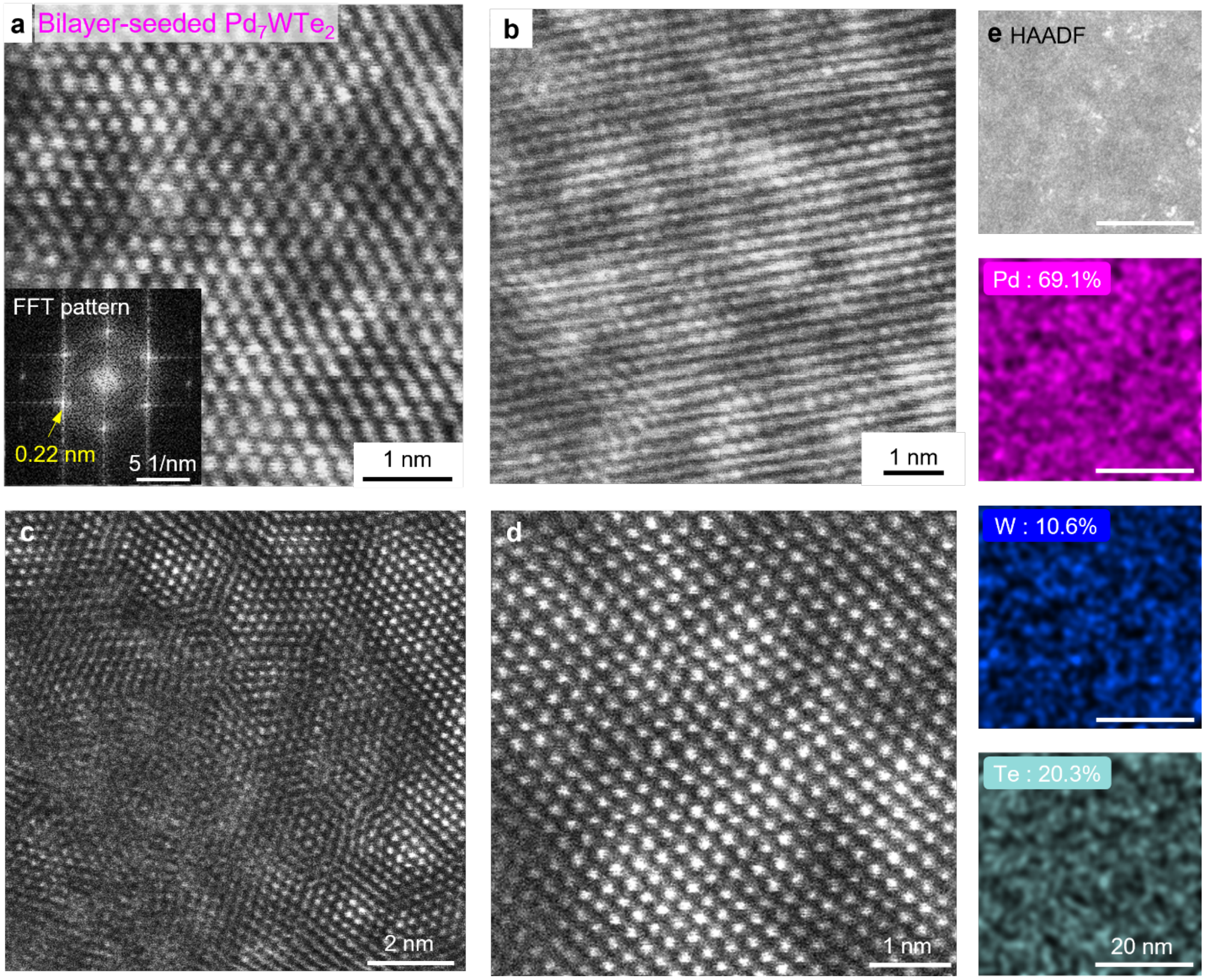

**Extended Data Fig. 11 | In-plane lattice structure and atomic fraction of bilayer-seeded $Pd_7WTe_2$. a,** An atomic-resolution STEM plan-view image of bilayer-seeded $Pd_7WTe_2$, which displays similar hexagonal lattice as monolayer seeded $Pd_7WTe_2$. The FFT pattern of such lattice is shown in the inset. **b-d,** Selected lattice structures observed in other locations of bilayer-seeded $Pd_7WTe_2$, indicating the presence of non-uniform domain structures. **e,** Elemental mappings and atomic fraction of bilayer-seeded $Pd_7WTe_2$.